# REBD: A Conceptual Framework for Big Data Requirements Engineering

Sandhya Rani Kourla, Eesha Putti and Mina Maleki

Department of Electrical and Computer Engineering and Computer Science,
University of Detroit Mercy, Detroit, MI, 48221, USA

## ABSTRACT

*Requirements engineering (RE), as a part of the project development life cycle, has increasingly been recognized as the key to ensuring on-time, on-budget, and goal-based delivery of software projects;compromising this vital phase is nothing but project failures. RE of big data projects is even more crucial because of the main characteristics of big data, including high volume, velocity, and variety. As the traditional RE methods and tools are user-centric rather than data-centric, employing these methodologies is insufficient to fulfill the RE processes for big data projects. Because of the importance of RE and limitations of traditional RE methodologies in the context of big data software projects, in this paper, a big data requirements engineering framework, named REBD, has been proposed. This conceptual framework describes the systematic plan to carry out big data projects starting from requirements engineering to the development, assuring successful execution, and increased productivity of the big data projects.*

## 1. INTRODUCTION

Requirements engineering (RE) is the first and most critical phase of the Software Development Life Cycle (SDLC). RE is the branch of engineering concerned with the real-world goals for, functions of, constraints on systems, and the relationship of these factors to precise specifications of system behavior [1-4]. Software requirements engineer (SRE) is responsible for translating stakeholders' needs, desires, and wishes of a software system into precise and formal software requirements specification. SREs need to communicate effectively and frequently with stakeholders to elicit, analyze, model, and manage their requirements. Doing this at the early stage of the software development helps ensure requirements are meeting stakeholders' needs while addressing compliance and staying on schedule and within budget [3].

In contrast, it has been indicated in many researches that performing improper and careless RE is one of the main sources of the time-consuming rework, inadequate deliveries, budget overruns, and consequently failures of the projects [3, 5-7]. The HMS Titanic (1912), the Mars Climate Orbiter (1999), the Apollo 13 (1970), Space Shuttle Challenger (1986), and Space Shuttle Columbia (2002) projects are some examples of high-profile projects broke down due to poor requirements engineering [7]. Also, according to PMI's Pulse of the profession (2018), 35% of companies proclaimed that the primary reason for their project failures is inaccurate requirements gathering [8]. As a consequence, performing a proper RE at the early stages of project development is critical in the success of projects, especially for big data projects.

In a newly released study, International Data Corporation (IDC) forecasts that the big data technology and analytics software market, which in 2018 reached $60.7 billion worldwide, is expected to grow at a five-year compound annual growth rate (CAGR) of 12.5% [9]. Big data is a technology that deals with data sets that are too large or complex to handle with traditional data processing techniques for capturing, storing, analyzing, searching, sharing, transferring, visualizing, querying, and updating of data. The main characteristics of big data technologies are 5V's: volume, velocity, variety, volatility, and variability [4, 10-12]. Volume refers to the massive amount of data; Velocity refers to the high growth rate of incoming data that needs to be processed and analyzed; Variety refers to many different forms of data; Volatility refers to the duration which data is valid and should be stored in the repository; Variability refers to data whose context changes invariably. For big data projects, it is essential to define the targets of the project at earlier stages. Time and cost reduction, finding insights from statistics, and optimizing selection making are the most common goals that make an organization to appreciate big data projects [4, 11]. RE phase in big data projects' development helps in achieving these goals and finding the business values associated with projects; These values help stakeholders to understand the importance of the project and its value in the market [13].

However, employing traditional RE activities, including gathering, analyzing, modeling, validating, and documenting requirements, is not quite efficient for big data projects. The main reason is that traditional RE methods are user-centric and deals with requirements apparent to users, while big data RE methodologies should be data-centric as well. It means there are lots of potential information, hidden patterns, and knowledge that can be extracted and discovered from a large amount of historical and actual data existing in big data projects. Moreover, RE activities in big data should isolate the requirements for infrastructures, analytical tools and techniques, and end-user applications using big data [14]. This isolation is needed because gathering requirements and defining all 5V's characteristics for end-user applications is challenging [12, 15], and SREs are unfamiliar about performing this task.

In this paper, we have proposed a big data requirements engineering framework, named REBD because of the importance of RE in the software project's success and the limitation of traditional RE activities for big data projects. This framework explains the detailed steps for carrying out RE activities on different software projects development for improving their success rate. This framework first identifies the project type and then performs corresponding RE processes on it.The proposed framework also helps to eradicate many challenges regarding knowledge discovery and balancing both data infrastructure and software development (SD) in big data projects.

The rest of the paper is organized as follows. After reviewing related works in section 2, section 3 describes traditional requirements engineering activities. Section 4 briefs big data requirements engineering; Section 5 explains the proposed REBD framework, and section 5 wraps up the research.

## 2. RELATED WORK

Requirements engineering in the context of big data applications is a hot research topic that attracted the attention of researchers in recent years. An analysis of the state of the art of big data RE research studies shows that little research has been conducted in this area by 2018 [11]. The investigation areas of this research included the phases of the RE process, type of requirements, application domains, RE research challenges, and solution proposals intended by RE research in the context of big data applications. In the following, some of the related work for big data RE will be presented.

To understand the context of the problem for the big data software applications, an empirically derived RE artifact model has been proposed in [15], which is equivalent to domain models. The proposed model can capture the fundamental RE components and the relationships concerning the development of big data software applications. "How can the requirements of a project be identified and analyzed to determine the suitability of a combined big data technology application?". To answer this question, a new process model has been proposed, detailed, and evaluated in [1]. This model considers the compound requirements, Knowledge Discovery in Databases (KDD), and a big data classification framework to identify the most relevant requirements of big data projects. Identified requirements for building big data applications must address big data characteristics (5V's) in terms of quality requirements. Quality requirements, also known as quality attributes, are those functional or nonfunctional requirements that used to measure the system's performance, such as reliability and availability. A new approach is proposed in [10] to ensure that the big data characteristics have adequately addressed in the specification of quality requirements.

Even though there are many requirements elicitation techniques in traditional RE, the researchers believe that more efficient tools and techniques should be employed to identify all requirements, business values, and knowledge from big data. Machine learning, deep learning, natural language processing, and data mining are some data analysis technologies that can use to discover requirements and valuable knowledge from big data [16-18]. Also, actionable use case (AUC) diagram is one of the efficient tools introduced in [13] that allows identified business values unleashed from data to be depicted in the data scientist's model, together with their roles in the software and their interactions with other software components. The AUC diagrams enhance the users' experience, optimize the systems' utility, and consequently maximize profit. Process mining, described in [19], is another efficient method that helps SRE to elicit, prioritize, and validate requirements from big data using execution logs, process discovery, and conformance techniques. The capability of process mining to discover valuable insights from event logs and processes of the system helps SRE to eradicate many challenges of traditional RE.

## 3. TRADITIONAL REQUIREMENTS ENGINEERING

As mentioned earlier, RE is one of the essential phases of the project's life cycle, and the failure of the RE leads to the failure of the project. So, software requirement engineers (SRE) are responsible for conducting the RE activities in a very well-mannered way, which can resolve many conflicts between stakeholders as well as among requirements. The main activities of traditional RE are listed below [2-3,13]:

**A. Requirements Elicitation:** Requirement elicitation as a critical activity in the requirement development process is the practice of uncovering and gathering the stakeholders' needs and desires from the system in terms of functional and nonfunctional requirements. To have a productive elicitation and dig all of the stakeholders' requirements, SRE should be able to communicate effectively with all stakeholders. Multiple techniques such as brainstorming, card sorting, laddering, interviews, prototyping, domain analysis, Joint Application Development (JAD), and Quality Function Deployment (QFD) can be employed to conduct requirements elicitation.

**B. Requirements Analysis:** Requirement analysis is the practice of defining the system boundaries and analyzing elicited and documented requirements to make sure that they are clear, concise, understandable, unambiguous, consistent, and complete. Also, requirements agreement, which is a process of resolving the conflicts in the requirements derived from different sources, is part of this activity. Use cases and user stories are often useful tools for this purpose.

**C. Requirements Modeling:** The requirement modeling intends to mold raw requirements into technical representations of the system. This activity guides SRE to archive all requirements during the specification process. Proper requirements representation eases the communication of requirements and conversion of those requirements into system architecture and design. To represent the analyzed functional and nonfunctional requirements in any systems, various modeling techniques can be employed, such as use cases, user stories, natural languages, formal languages, and a variety of formal, semiformal, and informal diagrams [3].

**D. Requirements Validation:** Requirements validation is the practice of checking and revisingrequirements to ensure that the specified requirements meet customer needs. Requirements validation also involves checking that (a) the system does not provide the functions that the customer does not need, (b) there are no requirements conflicts, and (c) time and budget constraints will meet. Systematic manual analysis of the requirements, test-case generation, comparative product evaluation tools, or some of the requirements elicitation techniques can be used for requirements validation [3].

**E. Requirements Specification:** It is the process of documenting the stakeholders' needs in terms of precise and formal software requirements specification. A software requirement specification (SRS) is a document that uses as a foundation of software development and acts as a contract between the software stakeholders and the developers. As a consequence, preparing an accurate SRS report at the end of the RE phase is critical in the success of projects. SRS contains the mission, scope, goals of the project, software and hardware requirements of the project, as well as functional and nonfunctional system requirements in terms of formal, semiformal, and formal models created in the previous processes.

## 4. BIG DATA REQUIRMENTS ENGINEERING

As mentioned earlier, the traditional RE methods are user-centric; focus on requirements apparent to users. However, big data RE methodologies should be data-centric as well because there are lots of potential information and hidden patterns that can be extracted from data by data scientists. As a consequence, big data RE model proposed in [13] consists of three different types of processes: processes drove by software requirements engineer (SRE), processes drove by data scientists (DS), and processes drove jointly by software requirements engineers and data scientists (SRE/DS). Table 1 demonstrates the RE activities for the big data products, including the RE activities, the responsible person for executing each activity, and their artifacts. Moreover, the column "Included in" indicates whether each activity is carried out in the traditional RE model (TRE) and/or big data RE model (BDRE).

From the table, it is clear that compared to the traditional RE, big data RE contains a few extra steps, including:

- **Data Acquisition:** This activity can be explained as collecting, gauzing, and fine-tuning the data before storing it in the data repository [13, 20]. The traditional collecting methods cannot make data acquisition because of the characteristics of big data and the high cost associated with it. Hence, data scientists use different data acquisition methods to discard useless data and keep important ones [21].

- **Data Analysis and Value Discovery:** In this activity, the acquired big data is analyzed by data scientists to reveal information and discover business values. Since data analysis is a time taking task and failing in better analysis will lead to consider a wrong decision, different data analysis technologies such as machine learning, deep learning, natural language processing, data mining, text analytics and predictive analytics help in fixing this issue [13, 16-18]. Moreover, techniques like process mining [19] and tools like smart grids [22] can be employed to accelerate the process of knowledge discovery from big data.

- **Use Case Consolidation:** Consolidation is merging the work and building models by software requirements engineers and data scientists. This consolidation is represented as an actionable use case diagram. The AUC diagram for big data software proposed in [13] is a merging of the data scientist model and the traditional use case. Once the values are defined and presented, the consolidated model is ready to be presented and discussed with the customers.

In addition to the requirements matrices, the big data SRS report contains business values extracted from data and AUC diagrams. Moreover, to ensure that big data characteristics are appropriately addressed, SRS should contain the quality attribute matrices. The quality attribute matrix is designed by intersecting a big data characteristic with a quality attribute and then identifying the system's quality requirements that apply to that intersection during big data RE process, as explained in [10].

## 5. CONCEPTUAL REBD FRAMEWORK

In this section, we describe our proposed big data requirements engineering framework, REBD, as depicted in Figure 1. This framework explains the planned approach for carrying out big data projects for improving the success rate. The purpose of this conceptual framework is to eradicate challenges like deciding whether to perform big data requirements engineering or traditional requirements engineering, knowledge discovery, and balancing the efforts to have the successful execution as described in the following.

Table 1. A Requirement Engineering Model for Big Data Software

| Process | Performed by | Output | Included in |
| --- | --- | --- | --- |
| Requirements Elicitation | SRE/DS along with Customers | List of requirements | TRE and BDRE |
| Data Acquisition | DS | Filtered data used for knowledge discovery | BDRE |
| Requirements Analysis | SRE | Technical specifications and SRE models | TRE and BDRE |
| Data Analysis | SRE | Extracted values and DS model | BDRE |
| Use case Consolidation | SRE /DS | Combined SRE and DS model | BDRE |
| Requirements Modelling | SRE | Finalized AUC diagrams | TRE and BDRE |
| Requirements Validation | SRE | Validated requirements | TRE and BDRE |
| Requirements Specification | SRE | SRS report | TRE and BDRE |

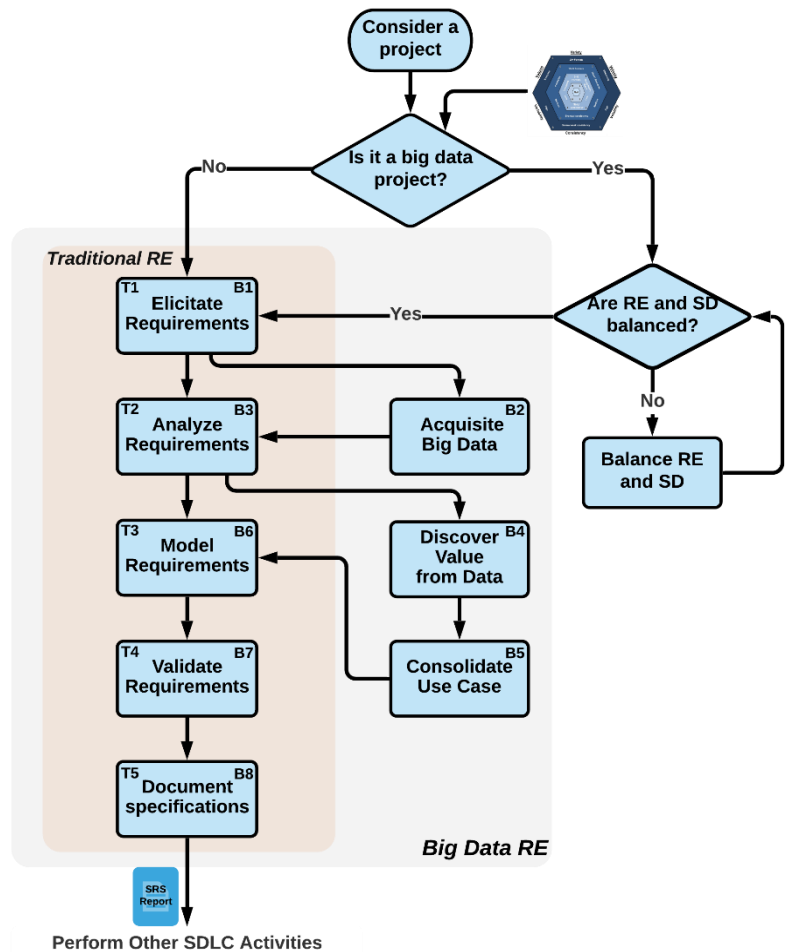

Figure 1. Requirements Engineering for Big Data (REBD) framework

Initially, when the project starts to develop, it is crucial to derive and analyze its characteristics to find whether the project falls under the category of big data or traditional projects. Doing this at the earliest phase of the project will save a considerable amount of time spent on the project lifecycle. To identify the category of the project, we have to figure out whether the recognized characteristics are big data characteristics or not. As mentioned earlier, the main characteristics of big data projects are 5V's: volume, velocity, variety, volatility, and variability. Next, a comprehensive quantitative classification model proposed by [1] is used to verify the project type. This model combines the recognized big data characteristics to figure out the project type. As shown in Figure 2, this classification model contains five layers in which the top four layers are willful to give an exact value depends on the severity, while the fifth level (NULL) is considered for characteristics that cannot be determined presently. If one of the supporting characteristics is assigned to this level, these will not be further utilized. Finally, all assigned values are added up and divided by the number of the addressed characteristics not assigned to the NULL layer. If the calculated assessment value is greater than or equal to 1.33, an application considers being a big data project.

If the project considers being in the group of traditional projects, traditional RE activities should carry out in the order specified in Figure 1 by T1 to T5 labels. However, if the project is found to be in the group of big data projects, then it should be checked whether RE and SD is balanced or not.

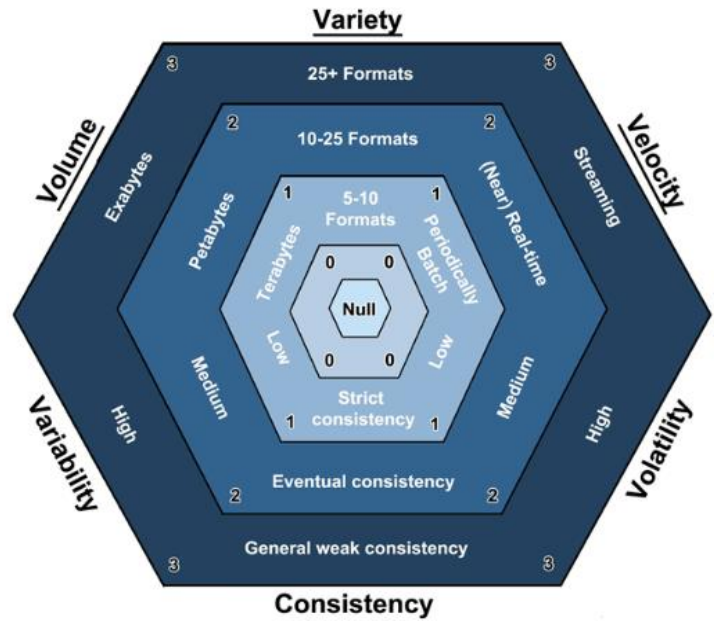

Figure 2. Big data characteristics and classification model [1]

Balancing RE and SD determines the challenges related to RE and SD activities. Challenges related to RE is identifying 5V's, which has been done already, and the challenges associated with SD are designing, coding, testing, and maintaining a big data end-user application. SD challenges are very much concerned because it may lead to the failure of the project, even after spending a lot of time in RE activities. To resolve this issue, a model has been proposed by [14], which contains a separate unit for doing the research needed for RE and SD. Obviously, anything without proof of concepts (PoC) and research may cause a shortage of time. The research and PoC, which give the optimal solution, should be used in the project practice. This ensures the balance between RE and SD, and the development of novel and better big data end-user applications. Once there is a solution for all challenges, then big data RE activities should carry out in order specified in Figure 1 by B1 to B8 labels.

After gathering engineered requirements documented in the SRS report, other processes of the software development life cycles, including design, implementation, testing, deployment, and maintenance, will start to produce software with the highest quality and lowest cost in the shortest time possible.

## 6. CONCLUSION

One of the main reasons behind the failure of many projects is improper requirements engineering and failing to capture the stakeholders' requirements, which leads to time-consuming rework, inadequate deliveries, and budget overruns. So, performing requirements engineering is a crucial phase of the project's development lifecycle. As big data is one of the new and emerging technology, developing big data projects and improving their success rates is an excellent achievement for companies. However, employing traditional RE activities is not sufficient to guarantee big data projects' development success because of the 5V's characteristics of big data.There is a lot of knowledge, valuable information, business values, and quality features hidden in a large amount of historical and actual raw data existing in big data projects. To analyze thesedata and discover hidden requirements, data scientists need to collaborate with SRE in big data RE activities.

In this paper, a big data requirements engineering framework, called REBD, has been introduced by considering the importance of RE as well as limitations of traditional RE methods. This framework explains the detailed steps for carrying out RE activities on different software projects and can be used to increase the success rate of project development. The proposed framework also considers the importance of balancing RE and SD in big data projects.

Although the current proposed framework is assumed to eradicate the challenges linked with big data, RE and SD, it should be tested practically for the project development to validate its efficiency. This research will be extended to discover better tools and techniques for knowledge discovery from big data. Also, the management of big data quality requirements in a preferred way will be investigated in the future.

**REFERENCES**

[1] M. Volk, N. Jamous, and K. Turowski, "Ask the right questions: requirements engineering for the execution of big data projects," presented at the 23rd Americas Conference on Information Systems, SIGITPROJMGMT, Boston, MA, Aug. 2017, pp 1-10.

[2] J. Dick, E. Hull, and K. Jackson, *Requirements Engineering*, 4th edition, Springer International Publishing, 2017.

[3] P.A. Laplante, *Requirements Engineering for Software and Systems,* 3rd Edition. Auerbach Publications (T&F), 20171024. VitalBook file, 2017.

[4] D. Arruda, "Requirements engineering in the context of big data applications," ACM SIGSOFT Software Engineering Notes, 43(1): 1-6, Mar. 2018.

[5] A.G. Khan, et. al., "Does software requirement elicitation and personality make any relation?" Journal of Advanced Research in Dynamical and Control Systems. 11. 1162-1168, 2019.

[6] A. Hussain, E. Mkpojiogu, and E. Kamal, "The role of requirements in the success or failure of software projects," presented at the International Soft Science Conference (ISSC), Malaysia, 2016, pp 6-7.

[7] T.A. Bahill and S.J. Henderson, "Requirements development, verification, and validation exhibited in famous failures," Systems Engineering, 8(1): 1–14, 2005.

[8] Pulse of the Profession 2018: Success in Disruptive Times, 2018.

[9] C. Gopal, et al., "Worldwide big data and analytics software forecast, 2019–2023," IDC Market Analysis, US44803719, Sept. 2019.

[10] I. Noorwali, D. Arruda, and N. H. Madhavji, "Understanding quality requirements in the context of big data systems," presented at the 2nd International Workshop on Big Data Software Engineering (BIGDSE), Austin, USA, May 2016, pp. 76-79.

[11] D. Arruda and N.H. Madhavji, "State of requirements engineering research in the context of big data applications," presented at the 24th International Working Conference on Requirements Engineering: Foundation for Software Quality (REFSQ), Utrecht, The Netherlands, Mar. 2018, pp 307-323.

[12] M. Volk, D. Staegemann, M. Pohl, and K. Turowski, "Challenging big data engineering: positioning of current and future development," presented at the 4th International Conference on Internet of Things, Big Data and Security (IoTBDS), Heraklion, Greece, May. 2019, pp 351-358.

[13] H.H. Altarturi, K. Ng, M.I.H. Ninggal, A.S.A. Nazri, and A.A.A. Ghani, "A requirement engineering model for big data software," presented at the 2nd International Conference on Big Data Analysis (ICBDA), Beijing, China, Mar. 2017, pp 111-117.

[14] NH. Madhavji, A. Miranskyy, and K. Kontogiannis, "Big picture of big data software Engineering: with example research challenges," presented at the IEEE/ACM 1st International Workshop on Big Data Software Engineering (BIGDSE), Florence, Italy, May 2015, pp 11-14.

[15] D. Arruda, N.H. Madhavji, and I. Noorwali, "A validation study of a requirements engineering artefact model for big data software development projects," presented at the 14th International Conference on Software Technologies (ICSOFT), Prague, Czech Republic, Jul. 2019, pp 106-116.

[16] B. Jan, et al., "Deep learning in big data analytics: a comparative study," Journal of Computer Electrical Engineering, 75(1): 275-287, 2019.

[17] A. Haldorai, A. Ramum, and C. Chow, "Editorial: big data innovation for sustainable cognitive computing," Mobile Netw Application Journal, 24(1): 221-226, 2019.

[18] R.H. Hariri, E.M. Fredericks, and K.M. Bowers, "Uncertainty in big data analytics: survey, opportunities, and challenges," Journal of Big Data, 6(1), 2019.
[19] M. Ghasemi, "What requirements engineering can learn from process mining," presented at the 1st International Workshop on Learning from other Disciplines for Requirements Engineering (D4RE), Banff, Canada, Aug. 2018, pp 8-11.
[20] K. Lyko, M. Nitzschke, and AC. Ngonga Ngomo, "Big data acquisition," in New Horizons for a Data-Driven Economy, Springer, Cham, 2016, pp 35-61.
[21] Z. Liu, P. Yang, and L. Zhang, "A sketch of big data technologies," presented at the 7th International Conference on Internet Computing for Engineering and Science (ICICSE), Shanghai, China, Sep. 2013, pp 26-29.
[22] X. Han, X. Wang, and H. Fan, "Requirements analysis and application research of big data in power network dispatching and planning," presented at the 3rd Information Technology and Mechatronics Engineering Conference (ITOEC), Chongqing Shi, China, Oct. 2017, pp 663-668.